# Direct Information Reweighted by Contact Templates: Improved RNA Contact Prediction by Combining Structural Features

Yiren Jian[2], Chen Zeng[2*], Yunjie Zhao[1*]

Author affiliation:

[1]Institute of Biophysics and Department of Physics, Central China Normal University, Wuhan 430079, China

[2]Department of Physics, The George Washington University, Washington, DC 20052, USA

*Corresponding author

Chen Zeng     Email: chenz@gwu.edu

Yunjie Zhao   Email: yjzhaowh@mail.ccnu.edu.cn

## Abstract

It is acknowledged that co-evolutionary nucleotide-nucleotide interactions are essential for RNA structures and functions. Currently, direct coupling analysis (DCA) infers nucleotide contacts in a sequence from its homologous sequence alignment across different species. DCA and similar approaches that use sequence information alone usually yield a low accuracy, especially when the available homologous sequences are limited. Here we present a new method that incorporates a Restricted Boltzmann Machine (RBM) to augment the information on sequence co-variations with structural patterns in contact inference. We thus name our method DIRECT that stands for Direct Information REweighted by Contact Templates. Benchmark tests demonstrate that DIRECT produces a substantial enhancement of 13% in accuracy on average for contact prediction in comparison to the traditional DCA. These results suggest that DIRECT could be used for improving predictions of RNA tertiary structures and functions. The source codes and dataset of DIRECT are available at http:// http://zhao.phy.ccnu.edu.cn:8122/DIRECT/index.html.

# Introduction

RNA molecules play critical roles in various biological processes (1-8). There remain various challenges in determination of RNA structure through experiments to get a comprehensive understanding of structure-function relation for RNAs (9). Many computational prediction methods for RNA tertiary structure have been developed, including homology or fragments based prediction (ModeRNA, Vfold, RNAComposer, 3dRNA) (10-15) and simulation-based prediction (SimRNA, Rosetta FARFAR, iFoldRNA, NAST) (16-20). Although the prediction accuracy of RNA tertiary structure has been improved over the years, the prediction task remains challenging for RNAs of complex topology. Therefore, information on the long-range contacts in an RNA, which dictates its topology, is needed for its structure and function characterization.

The knowledge of nucleotide-nucleotide interaction provides the distance restraint information that can be used to improve RNA structure prediction. One of the most successful methods for contact prediction is the direct coupling analysis (DCA). DCA infers the interacting nucleotides in a sequence from the sequence coevolution across different species (21-32). A recent mean-field formulation of DCA (mfDCA) provides an efficient computational framework to extract the direct contact information and has been applied to many RNAs. It has been shown that DCA provides sufficient native intra-domain and inter-domain nucleotide-nucleotide contact information for riboswitch and RNA-protein complexes (33-35). In addition to DCA, there are also network or machine learning based approaches to infer covariation signals from multiple sequence alignments (36-40). The common feature among these approaches is the exclusive usage of evolutionary information extracted from homologous sequences. The prediction accuracy thus depends on accurate multiple sequence alignments of thousand or more homologous sequences.

There is also a structure-based approach to improve the contact prediction from sequence variations alone. Skwark *et al.* applied a pattern-recognition approach to the contact prediction of a pair of residues by examining the expected pattern of nearby contacts surrounding the pair (41). Specifically, a 3x3 matrix of local contacts is constructed as follows. Each residue of the pair is expanded into a fragment of three residues by including the two neighbors, and all residue-residue contacts between the two fragments form the 3x3 matrix with element value of 1 for contact and 0 for non-contact. It was found that a contact at the center of the 3x3 matrix is typically surrounded by three other contacts in the matrix and a non-contact at the center, however, is likely surrounded by no more than one other contact. By incorporating these local contact patterns, this approach is able to improve the secondary structures of alpha helices and beta strands for protein structures.

However, it is more important and difficult to pinpoint loop-loop interactions in an RNA structure than to identify its secondary structure of base-pair interactions. This may require some global information beyond the local pattern modeled by a small 3x3 contact matrix. We therefore introduce a new method to improve RNA contact prediction that combines global structure information learned by a Restricted Boltzmann Machine (RBM) and sequence co-evolution information obtained by DCA. We thus named our method Direct Information REweighted by Contact Templates (DIRECT). In a benchmark testing on riboswitch, DIRECT outperforms the state-of-the-art DCA predictions for long-range contacts and loop-loop contacts. Moreover, DIRECT maintains better predictions when we randomly select only 50 homologous sequences to mimic the situation of insufficient sequences for the 5 RNAs tested with the 50 sequences representing from 11% to 43% of available sequences. The $6^{th}$ RNA (3OWI) in excluded because the number of its available sequences is already less than 50.

## Methods

### *Inference workflow*

DIRECT (Direct Information REweighted by Contact Templates) improves prediction of tertiary contacts by using both sequence and structure information. Figure 1 illustrates the workflow of DIRECT. The inference framework consists of completely hierarchical modules and thus offers the flexibility to incorporate more sequences and structures that may become available in the future as well as further improved DCA for enhanced performance.

### *Restricted Boltzmann Machine (RBM)*

The Restricted Boltzmann Machine (RBM) is a graphical model for unsupervised learning that can extract features from the input data (42). RBM has a visible layer and a hidden layer. The restriction is that units in the visible layer only interact with units from the hidden layer. This network structure leads to a factorized probability for observing a give configuration, which in turn further simplifies the learning process. The energy of a RBM is given by

$$E(v, h|W, b, c) = -b^T v - c^T h - h^T W v \tag{1}$$

where W is the connection weight matrix between visible and hidden units. b, c are bias units as offsets. The probability of having a given v, h is then

$$p(v, h|W, b, c) = \frac{1}{z(W,b,c)} e^{-E(v,h|W,b,c)} \tag{2}$$

$$z(W, b, c) = \sum_{v,h} e^{-E(v,h|W,b,c)} \tag{3}$$

where $z(W, b, c)$ is the partition function that sums up all possible v and h. The RBM is trained through stochastic gradient descent (SGD) on negative log-likelihood of the empirical data. $L(W, c, b, T)$ is defined as the loss function:

$$L(W, c, b, T) = -\frac{1}{N} \sum_{v \in T} \log P(v|W, b, c) \tag{4}$$

where $P(v|W, b, c)$ is given by

$$P(v|W, b, c) = \sum_h p(v, h|W, b, c) \tag{5}$$

T above is a set of samples from the empirical data. By minimizing the loss function, we can update the parameters W, b, c according to the equations below:

$$W = W - \frac{\partial L(W,b,c,T)}{\partial W} \tag{6}$$

$$b = b - \frac{\partial L(W,b,c,T)}{\partial b} \tag{7}$$

$$c = c - \frac{\partial L(W,b,c,T)}{\partial c} \tag{8}$$

*Contact definition and evaluation criteria*

Two nucleotides are considered in contact if they contain a pair of heavy atoms, one from each nucleotide, less than a pre-defined cutoff (43,44). To compare DIRECT with other earlier methods, we use the same contact distance cutoff of 8 Å as in previous studies (33,34). Since adjacent nucleotides in a sequence are always in contact, we only consider contacts between nucleotides that are separated by more than four nucleotides in a sequence to measure tertiary contacts of interest. To evaluate the quality of a prediction, we compute the positive predictive value (PPV) as follows.

$$PPV = \frac{|TP|}{|TP|+|FP|} \tag{9}$$

where TP (FP) denotes the true (false) positive and |TP| (|FP|) stands for the number of true (false) positives.

## *Training and testing sets*

Riboswitch is a regulatory portion of a messenger RNA. When binding with a small ligand, this regulatory segment will regulate the translation of the entire mRNA. In this study, we will focus on regulatory RNA such as riboswitch. To generate a representative set of riboswitch families for our study, we systematically select riboswitch families from the Rfam database. The ten representative riboswitches in the training set are shown in Table 1. For the testing set, we use the published testing dataset including six riboswitches (Table 2) (33).

## *Weight of structural information learned by RBM for prediction of riboswitch*

The Restricted Boltzmann Machine (RBM) is used to extract the contact knowledge from riboswitch structures in the training set (Figure 2).

Step 1: Generating one-dimensional input. Riboswitch structures in the training set are converted into contact maps by applying the distance cutoff of 8 Å. The lengths of the testing riboswitches range from 52 to 94 nucleotides. For the convenience of integrating the templates of structural information, all distance maps are resized by linear interpolation into the same size of 100x100 before applying the distance cutoff of 8 Å. To remove the overlap between the training set and testing set, we exclude all homologous training structures with respect to the riboswitch structure in the testing set for each prediction. To be more precise for this blind test, when predicting each of the six riboswitches in the test set, the targeting riboswitch and all its homologs are removed from the training set resulting only 9 RNA riboswitch structures in the training set. There are six different weights of structural information learned by RBM for each of the six riboswitches. We convert the lower triangle contact maps into a one-dimensional array with one channel per contact (as 1) or non-contact (as 0).

Step 2: Learning the shared weights. Training the RBM efficiently by stochastic gradient descent (SGD) involves an algorithm called Contrastive-Divergence (CD) invented by Hinton (45). In this study, we use a typical learning rate of 0.1 and epochs of 10000 during RBM training.

Step 3: Counting RBM-based structural contact frequency. After RBM is trained from the structures of existing riboswitch RNA, we generate 10,000 new structures and keep the last 5,000 structures to model the equilibrium that represents RBM's belief for the most common structure of riboswitches. We count the contact frequency for each nucleotide among these 5,000 structures and take this frequency as the final weight matrix learned by RBM on the structure information of the riboswitch.

*Direct coupling analysis*

The direct coupling analysis (DCA) is performed to infer the interacting nucleotides from sequence coevolution across different species (21,46-48). We first remove the sequences with gaps more than 50% in multiple sequence alignment (MSA) and then calculate the amino acid frequencies for single nucleotide and a pair of nucleotides. The direct couplings are defined as

$$DI_{ij} = \sum_{AB} P_{ij}^d(A,B) \ln \frac{P_{ij}^d(A,B)}{f_i(A)f_j(B)} \tag{10}$$

with the help of an isolated two-site model

$$P_{ij}^d(A,B) = exp\{e_{ij}(A,B) + \tilde{h}_i(A) + \tilde{h}_j(B)\}/Z_{ij} \tag{11}$$

$\tilde{h}_i(A)$ and $\tilde{h}_j(B)$ are defined by the empirical single-nucleotide frequency $f_i(A) = \sum_B P_{ij}^d(A,B)$ and $f_j(B) = \sum_A P_{ij}^d(A,B)$. See Morcos *et al.* (21) for details.

*Direct Information scores reweighted by structural contact frequency*

The final contact prediction is DI scores reweighted by structural information learned by RBM with better contact prediction accuracy.

$$DIRECT = DI \times W^2 \tag{12}$$

where DI is the direct information by direct coupling analysis, W is RBM-based structural contact frequency. Among the different powers of W considered (up to the 4[th] power), we finally select the 2[nd] power of W as in Eq. (12) to balance the contributions from both patterns of sequence evolution and RBM-based structural contact frequency.

## Results
**DIRECT achieves better overall performance**

Traditional direct coupling analysis (DCA) for RNA contact prediction has its own drawbacks. For one, DCA requires a sufficient number of homologous sequences for accurate sequence co-evolution analysis, which may not be readily available. Moreover, a co-evolving pair of nucleotides in an RNA molecule does not necessarily form an intra-molecular contact, but as an inter-molecular interaction across the homodimer interface of the RNA. Still many unknown factors other than intra- or inter-molecular interactions can result in co-evolving pairs and thus make it difficult to detect the true contacts among the evolving pairs without additional information. One way to overcome this difficulty is to augment the contact detection of a target RNA sequence with additional information on the structural contact template expected of the RNA class to which the target RNA belongs. To this end, we employ a Restricted Boltzmann Machine to learn the global contact template of any specific RNA class by using the known structures in that class. This contact template is then used to improve the contact prediction for new sequences in that class.

To evaluate our method DIRECT described in Methods, we use a riboswitch benchmark dataset (Table 2) published in Ref. 33. Six target RNAs are tested as shown in Figure 3. For a given target RNA, the RNA itself and its homologs are removed from the training set used to build the riboswitch structural contact template that will be combined with DCA to form DIRECT. We compare the success rate of DCA and DIRECT in predicting the true intra-molecular contacts from the top detected co-evolving pairs (up to top 100). As shown in Figure 3, an increase of 5%~7% in precision (positive predictive value defined in Methods) is clearly seen when DIRECT is compared to DCA for 1Y26, 2GDI, 2GIS, and 3IRW predictions. There is also a slight increase of 2% for 3OWI prediction. The improvement continues beyond the top 100 pairs. The only exception is 3VRS, for it differs from others by its higher-order RNA architecture stabilized by pseudoknots with few standard Watson-Crick pairs, which may lead to a low accuracy for contact prediction.

**DIRECT improves predictions for long-range contacts**

A contact range measures the sequence distance, i.e., the number of nucleotides (nt), between the two nucleotides in the contact. Contacts at different range convey different information. While short-range contacts in an RNA molecule reflect its local secondary structure, long-range contacts dictate the topology of its global structure. Prediction on long-range contacts remains difficult for most traditional methods. DCA predicts more accurately for short- (5~12nt) or medium-range (13~24nt) contacts, but less accurately for long-range (24nt+) contacts. DIRECT, however, utilizes the structural contact template learnt from the training set to re-rank DCA predictions and is able to improve the long-range contact prediction as shown in Table 3. Even a slight improvement in long-range contact prediction can have a significant impact on the accuracy and speed of RNA tertiary structure modeling because long-range contacts drastically reduce the structural space that needs to be searched for modeling.

**DIRECT captures more tertiary structural features**

The interaction types between different RNA secondary structure elements vary significantly. According to Chargaff's second parity rule, base-pair contacts are easier to predict. However, it remains difficult to predict long-range tertiary contacts. Since DIRECT is designed to capture the structural contact template independent of contact types in RNA, it should improve the prediction accuracy for all types of contacts. To verify this, we divide the tertiary contacts into four types of interactions, i.e., stem-loop, loop-loop, intra stem-stem, and inter stem-stem contacts. The intra stem-stem contacts between two nucleotides in the same stem determine the stem topology such as bending or twisting. On the other hand, contacts of stem-loop, loop-loop, and inter stem-stem can be used as distance constraints on the RNA tertiary fold.

In Table 4, it can be seen that the largest improvement of predictions by DIRECT lies in tertiary structural contacts. The correct prediction of base pairs can determine RNA secondary structure. The prediction accuracies of base pairs are similar between DCA and DIRECT. These results show that DCA already performs well for base pairs prediction. In contrast, contacts involving tertiary interactions are markedly improved by DIRECT. There are significant increases of 3~8 intra stem-stem contacts correctly predicted for 1Y26, 2GIS, 3OWI, and 3IRW, suggesting more bending or twisting contacts in these RNA structures. A more pronounced effect can be observed for other three types of contacts (loop-loop, loop-stem, and inter stem-stem) predictions. In particular, contacts involving loop regions are more accurately predicted. The results show that DIRECT predicts better tertiary fold.

**DIRECT identifies the riboswitch structure template**

Pattern recognition of RNA motifs would be a useful resource for RNA structure modeling and design since many structural motifs are highly conserved in homologous RNAs. While an expert curator could identify RNA motifs with ease, it is still difficult for a computer to recognize them automatically (49-51). We define the RBM-based contact template to be the top 100 pairs of weight learned by RBM. This contact template potentially offers an automated scheme for identification and classification of RNA structural motifs. To construct the contact template for riboswitch as an RNA class, a representative set of contact maps are extracted from non-redundant riboswitch structures and used as the training set to a Restricted Boltzmann Machine to learn the structural motifs in riboswitch.

To compare the contact template for the class of riboswitch learned by RBM to the experimental contact map of a specific target riboswitch, we resize the template to match the target and highlight their overlaps as shown in Figure 5 for six target riboswitches (1Y26, 2GDI, 2GIS, 3IRW, 3OWI, and 3VRS). It is noted that the target riboswitches share a high similarity in contact with the riboswitch template. The template captures roughly 12% of the experimental contacts. In contrast, the riboswitch template is very different from the experimental contact maps of non-riboswitch RNAs such as tRNA (1GTR) and ribozyme (2QUW) with low similarity as shown in Figure 6. Only 4% of experimental contacts of these two non-riboswitch RNAs overlap with the riboswitch template indicating the discriminative power of the template. It is thus not surprising that Figure 7 shows a significant decrease in contact prediction accuracy for tRNA (1GTR) and ribozyme (2QUW) by DIRECT using the riboswitch template. Taken together, these results suggest that riboswitch template captures the riboswitch structural motif pattern and recognizes riboswitch structures.

## Discussions

Previous research suggests the number of sequences should be more than three times the length of the molecule for reliable contact prediction (27). However, many RNA families do not satisfy this condition. While loosening the criterion for homology may results in more sequences, this approach inevitably leads to low accuracy in contact prediction. It remains challenging to extract evolutionary information from an insufficient number of sequences. To check if DIRECT can address this difficulty of insufficient sequences, we randomly select only 50 sequences to perform contact predictions on 5 target riboswitches whose lengths range from 52 to 92 nucleotides and already exceed the number of sequences used. The results given in Table 5 show that DIRECT outperforms DCA with an average increase of 12% in prediction precision suggesting that DIRECT can improve predictions even when the number of homologous sequences is insufficient.

In summary, we develop a hybrid approach that incorporates a Restricted Boltzmann Machine (RBM) to augment the information on sequence co-variations with structural templates in contact inference. Our results demonstrate a 13% precision increase for RNA contact prediction when structural templates are utilized. In fact, our approach establishes a straightforward framework that can incorporate any additional information such as NMR spectroscopy data by training a corresponding Restrictive Boltzmann Machine to further improve the prediction on RNA contacts.

## Acknowledgments

This work is supported by National Natural Science Foundation of China 11704140, Natural Science Foundation of Hubei 2017CFB116, the Thousand Talents Plan 31103201603, and self-determined research funds of CCNU from the colleges' basic research and operation of MOE 20205170045 to YZ.

## Author Contributions
YR performs most computational analysis under the supervision of CZ and YZ. YZ and CZ supervise the overall study and write the paper.

## Tables and Figures

**Table 1.** Selected riboswitches in the training set. The columns are PDB ID, experimental method, structural resolution, selected chain, and sequence length, respectively.

| PDB | Method | Resolution (Å) | Chain | Length |
|---|---|---|---|---|
| 2CKY | X-RAY | 2.90 | A | 77 |
| 2MIY | NMR |  | A | 59 |
| 3DJ2 | X-RAY | 2.50 | A | 174 |
| 3F2Q | X-RAY | 2.95 | X | 112 |
| 3IQR | X-RAY | 2.55 | A | 94 |
| 3MXH | X-RAY | 2.30 | R | 92 |
| 3OWW | X-RAY | 2.80 | A | 88 |
| 4EN5 | X-RAY | 2.96 | A | 52 |
| 4GMA | X-RAY | 3.94 | Z | 210 |
| 5C7U | X-RAY | 3.05 | B | 67 |

**Table 2.** Selected riboswitches in the testing set. The columns are PDB ID, riboswitch type, experimental method, structural resolution, selected chain, sequence length, and Rfam ID, respectively.

| PDB | Riboswitch type | Method | Resolution (Å) | Chain | Length | Rfam |
|---|---|---|---|---|---|---|
| 1Y26 | Adenine | X-RAY | 2.10 | X | 63 | RF00167 |
| 2GDI | TPP | X-RAY | 2.05 | X | 75 | RF00059 |
| 2GIS | SAM-I | X-RAY | 2.90 | A | 94 | RF00162 |
| 3IRW | c-di-GMP | X-RAY | 2.70 | R | 90 | RF01051 |
| 3OWI | Glycine | X-RAY | 2.85 | A | 86 | RF00504 |
| 3VRS | Fluoride | X-RAY | 2.60 | A | 52 | RF01734 |

**Table 3.** Number of correctly predicted contacts among the top 100 predictions further grouped into three categories according to their base-pair (bp) separation in sequence for short (5~12bps), medium (13~24bps), and long (>24bps) ranges, respectively. Results obtained from two different methods, DCA and DIRECT, are provided for comparison.

| PDB | Ranges | DCA | DIRECT |
|---|---|---|---|
| 1Y26 | 5~12 | 10 | 11 |
|  | 13~24 | 12 | 15 |
|  | 24+ | 8 | 9 |
| 2GDI | 5~12 | 10 | 13 |
|  | 13~24 | 6 | 5 |
|  | 24+ | 18 | 21 |
| 2GIS | 5~12 | 9 | 6 |
|  | 13~24 | 8 | 9 |
|  | 24+ | 15 | 20 |
| 3OWI | 5~12 | 4 | 3 |
|  | 13~24 | 5 | 1 |
|  | 24+ | 11 | 17 |
| 3IRW | 5~12 | 6 | 7 |
|  | 13~24 | 6 | 6 |
|  | 24+ | 21 | 25 |
| 3VRS | 5~12 | 8 | 9 |
|  | 13~24 | 1 | 1 |
|  | 24+ | 7 | 6 |

**Table 4.** Number of correctly predicted contacts among top 100 predictions for different classes of RNA. Results obtained from two different methods, DCA and DIRECT, are provided for comparison.

| PDB | Interactions | DCA | DIRECT |
|---|---|---|---|
| 1Y26 | Base pairs | 15 | 15 |
| | stem-loop | 4 | 4 |
| | loop-loop | 2 | 2 |
| | stem-stem(intra) | 6 | 10 |
| | stem-stem(inter) | 3 | 4 |
| 2GDI | Base pairs | 12 | 13 |
| | stem-loop | 12 | 14 |
| | loop-loop | 6 | 8 |
| | stem-stem(intra) | 3 | 3 |
| | stem-stem(inter) | 1 | 1 |
| 2GIS | Base pairs | 20 | 19 |
| | stem-loop | 1 | 3 |
| | loop-loop | 9 | 5 |
| | stem-stem(intra) | 2 | 7 |
| | stem-stem(inter) | 0 | 1 |
| 3OWI | Base pairs | 13 | 10 |
| | stem-loop | 3 | 1 |
| | loop-loop | 1 | 3 |
| | stem-stem(intra) | 3 | 6 |
| | stem-stem(inter) | 0 | 1 |
| 3IRW | Base pairs | 16 | 16 |
| | stem-loop | 6 | 7 |
| | loop-loop | 2 | 2 |
| | stem-stem(intra) | 8 | 12 |
| | stem-stem(inter) | 1 | 1 |
| 3VRS | Base pairs | 6 | 6 |
| | stem-loop | 1 | 1 |
| | loop-loop | 7 | 7 |
| | stem-stem(intra) | 2 | 2 |
| | stem-stem(inter) | 0 | 0 |

**Table 5.** Comparison for the number of correctly predicted contacts among the top 100 predictions of DCA and DIRECT in a simulated dataset with some sequences randomly removed. This examines the model performance when available sequences are insufficient.

| PDB | DCA | DIRECT |
|---|---|---|
| 1Y26 | 27 | 33 |
| 2GDI | 27 | 29 |
| 2GIS | 22 | 23 |
| 3OWI | 20 | 21 |
| 3IRW | 22 | 29 |
| 3VRS | 19 | 18 |

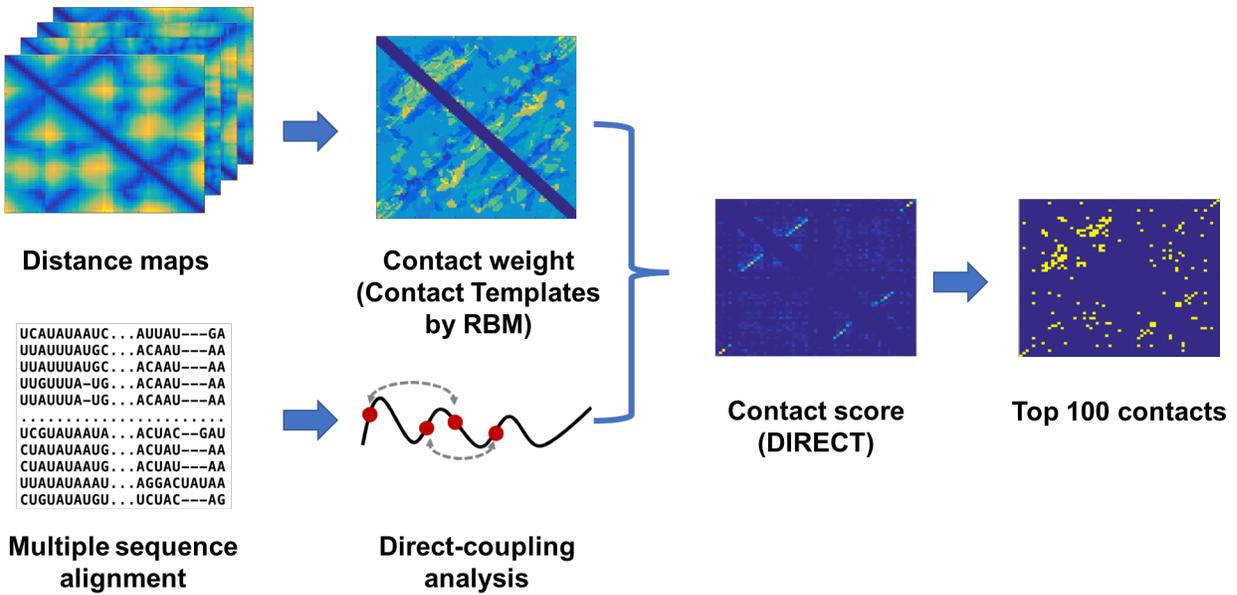

**Figure 1. Basic workflow of DIRECT for RNA tertiary contact prediction.** The corresponding RNA multiple sequence alignment (MSA) is extracted from Rfam database. The traditional direct-coupling analysis (DCA) predicts the tertiary contacts from sequence coevolution in MSA. DIRECT then reweighs the contacts by using structural templates trained by Restricted Boltzmann Machine (RBM). The reweighted contact prediction leads to a better overall performance.

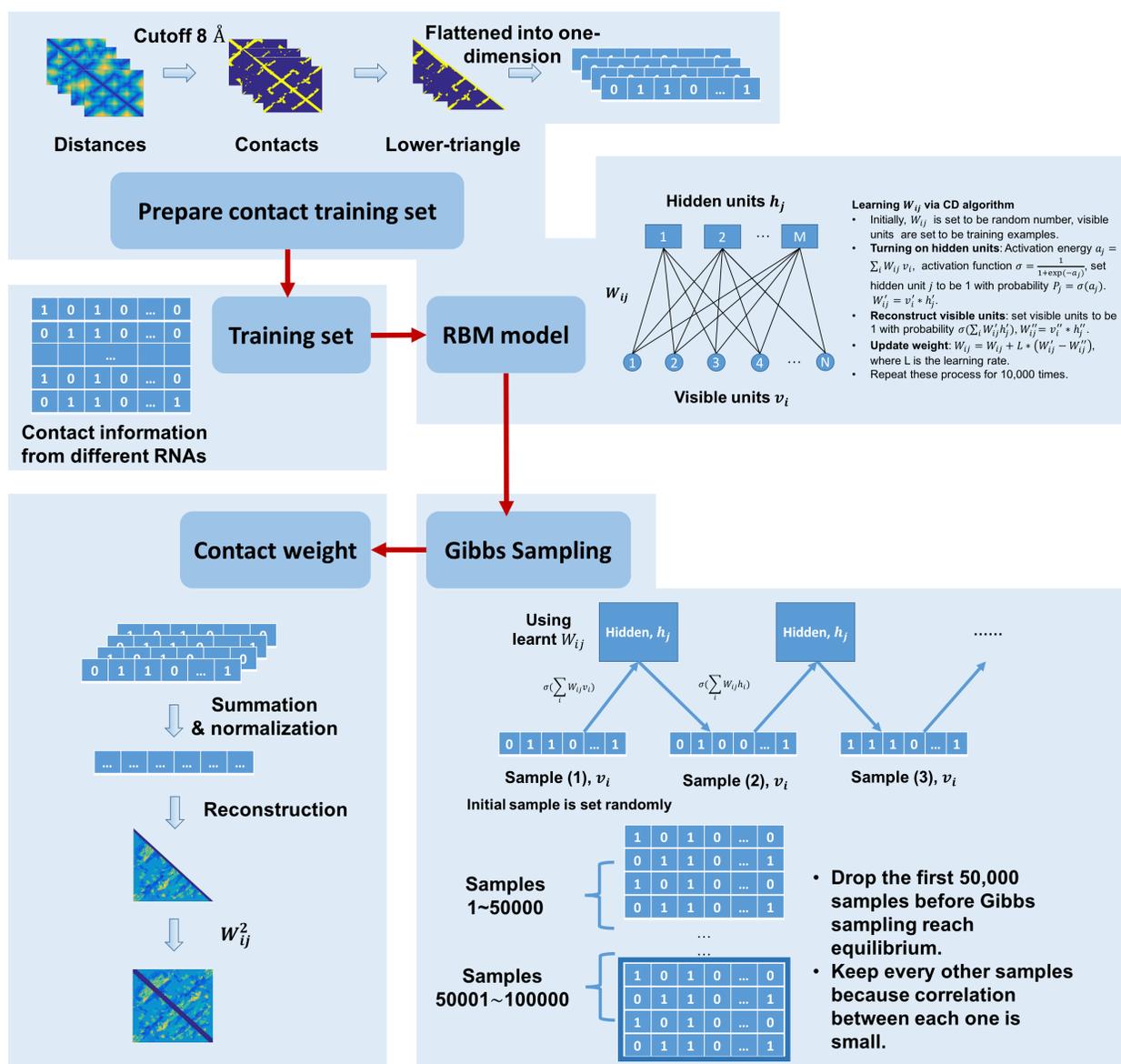

**Figure 2. Further refined workflow for part of Figure 1 on training a Restrictive Boltzmann Machine (RBM) to detect contact patterns.** Specific steps to extract the contact weights from RNA tertiary structure are as follows. (1) Prepare contact training set. A contact map of a given RNA is constructed from its nucleotide-nucleotide distance matrix. Two nucleotides are considered in contact if a pair of heavy atoms, one from each nucleotide, is less than 8 Å apart. The lower triangles of the contact map is maintained and then converted to a one-dimensional array as the input to RBM. (2) Training set. The training set consists of all contact maps of riboswitch structures but with the testing homologous riboswitch structure removed. (3) RBM model. Parameters in RBM are trained by the Contrastive Divergence (CD) algorithm. (4) Gibbs sampling. The Gibbs sampling method is used to generate new contact maps using RBM model. The last 50,000 samples are maintained for contact weight calculation. (5) Contact weight. The Gibbs sampling results are normalized into one contact matrix representing nucleotide-nucleotide contact weights for a typical riboswitch structure.

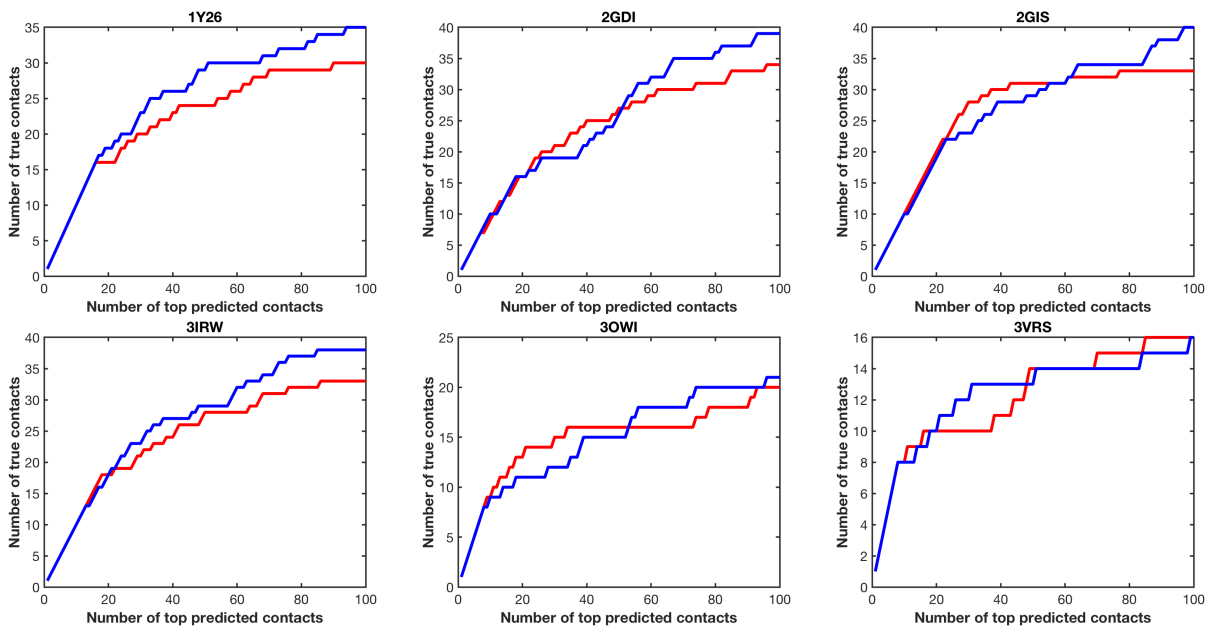

**Figure 3. Accuracy of nucleotide-nucleotide contact prediction for all six RNAs in the testing set.** The number of true contacts among a number of top predicted contacts is shown for each of the six RNAs. With the exception of 3VRS, DIRECT (blue lines) achieves 13% higher accuracy on average than DCA (red lines) for true contacts among the top 100 predicted contacts.

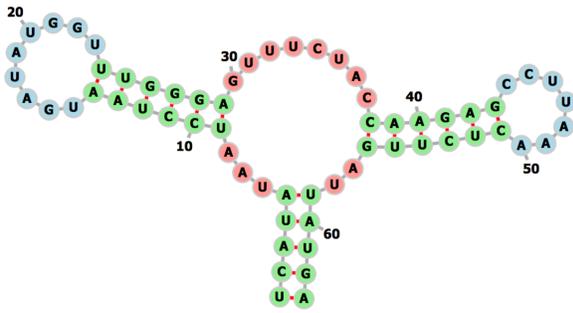
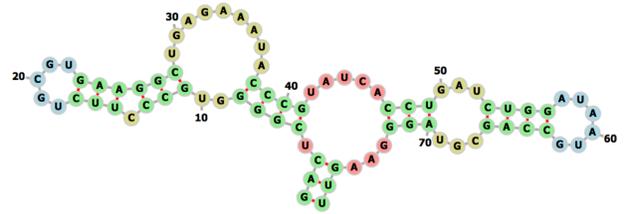

**1Y26 RF00167**  **2GDI RF00059**

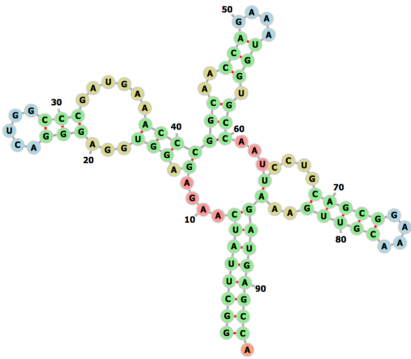
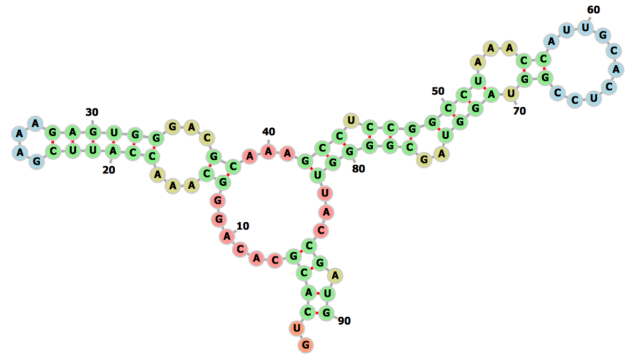

**2GIS RF00162**  **3IRW RF01051**

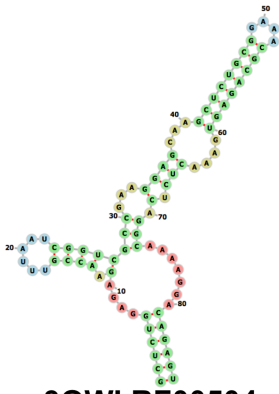
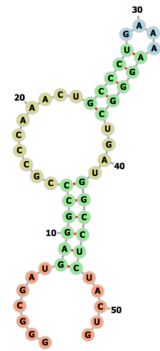

**3OWI RF00504**  **3VRS RF01734**

**Figure 4.** The secondary structures of RNAs in the testing set. The green nucleotides are for base pairs, blue for hairpin loops, yellow for internal/bulge loops, and red for junctions or exterior loop. The secondary structures are generated by Forna program.

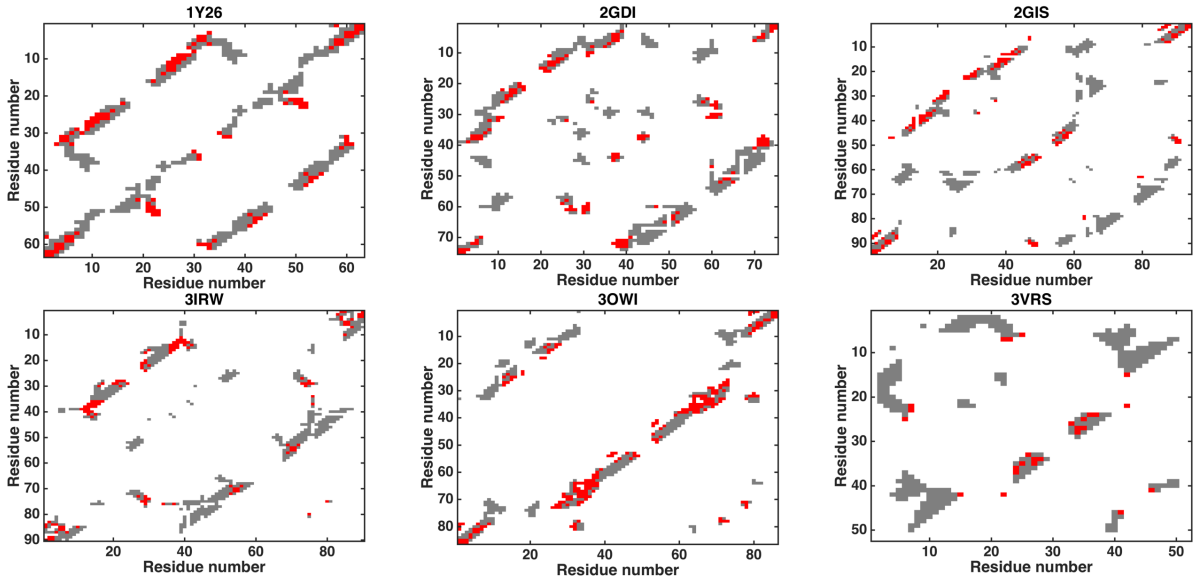

**Figure 5. Comparison of riboswitch contact maps produced by the true experimental structures and riboswitch-based RBM in blind tests.** Gray dots represent the true contacts between nucleotides in riboswitch and the red dots are the true contacts also correctly predicted by the riboswitch-based RBM. With the exception of 3VRS, the two contact maps shared an overall high similarity that can be used for riboswitch recognition by riboswitch-based RBM. 3VRS differs from others by its higher-order RNA architecture stabilized by pseudoknots with few standard Watson-Crick pairs, which may lead to a low accuracy of RBM prediction.

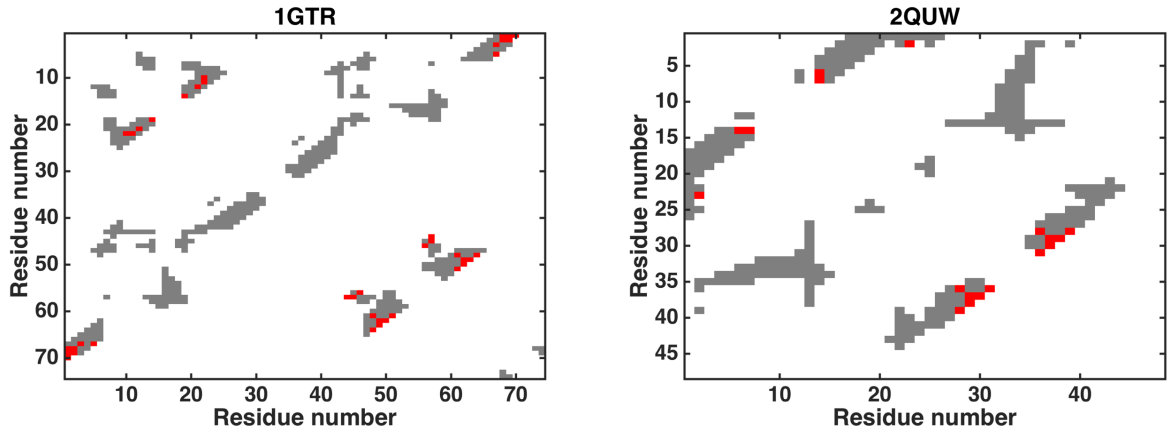

**Figure 6. High-ranking riboswitch-RBM-based contact map shares low similarity to RNAs that are not riboswitch.** Grey dots in the plots represent the true contacts between nucleotides in RNAs and red dots are the same patterns in both a RNA contact map and riboswitch-RBM-based contact map. The low similarity of the contact map shows riboswitch-RBM-based contact map is different from these RNAs.

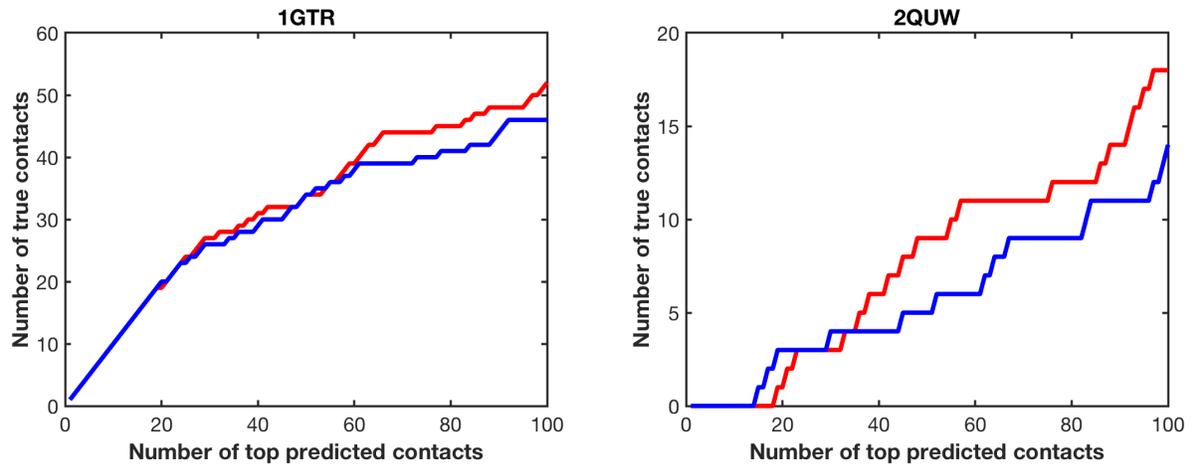

**Figure 7. Contact prediction accuracy for two non-riboswitch RNAs.** The precisions of nucleotide-nucleotide interaction predictions by DIRECT (blue lines), which incorporates riboswitch-based RBM, are consistently lower than those by the traditional DCA (red lines), leading to an average decrease of 17% in the top 100 predictions.